\pdfoutput=1
\documentclass[aps,PRB,twocolumn,amsmath,amssymb,superscriptaddress,showpacs]{revtex4-1}
\usepackage{graphicx}
\usepackage{dcolumn}
\usepackage{bm}
\usepackage{textcomp}
\usepackage{hyperref}

\begin{document}


\title{Multigap superconductivity in locally non-centrosymmetric SrPtAs: An $^{75}$As nuclear quadrupole resonance investigation}
\author{F. Br\"uckner} \email{fxbr.pc@gmail.com}
\author{R. Sarkar}
\author{M. G\"unther}
\affiliation{Institute for Solid State Physics, TU Dresden, D-01069 Dresden, Germany}
\author{H. K\"uhne}
\affiliation{National High Magnetic Field Laboratory, Florida State University, Tallahassee, Florida 32310, USA}
\affiliation{Dresden High Magnetic Field Laboratory (HLD), Helmholtz-Zentrum Dresden-Rossendorf, D-01314 Dresden, Germany}
\author{H. Luetkens}
\affiliation{Laboratory for Muon-Spin Spectroscopy, Paul Scherrer Institute, CH-5232 Villigen PSI, Switzerland}
\author{T. Neupert}
\affiliation{Princeton Center for Theoretical Science, Princeton University, Princeton, New Jersey 08544, USA}
\author{A. P. Reyes}
\affiliation{National High Magnetic Field Laboratory, Florida State University, Tallahassee, Florida 32310, USA}
\author{P. L. Kuhns}
\affiliation{National High Magnetic Field Laboratory, Florida State University, Tallahassee, Florida 32310, USA}
\author{P. K. Biswas}
\affiliation{Laboratory for Muon-Spin Spectroscopy, Paul Scherrer Institute, CH-5232 Villigen PSI, Switzerland}
\author{T. St\"urzer}
\author{D. Johrendt}
\affiliation{Department Chemie, Ludwig-Maximilians-Universit\"at M\"unchen, D-81377 Munich, Germany}
\author{H.-H. Klauss}
\affiliation{Institute for Solid State Physics, TU Dresden, D-01069 Dresden, Germany}

\date{\today}

\begin{abstract}
We report detailed $^{75}$As-NQR investigations of the locally non-centrosymmetric superconductor SrPtAs. The spin-lattice relaxation studies prove weakly coupled multi-gap superconductivity. The Hebel-Slichter peak, a hallmark of conventional superconductivity, is strongly suppressed, which points to an unconventional superconducting state. The observed behavior excludes a superconducting order parameter with line nodes and is consistent with proposed \emph{f}-wave and chiral \emph{d}-wave order parameters.
\end{abstract}

\maketitle

\section{Introduction}

After the discovery of superconductivity in doped transition metal pnictides, which is often referred to as a milestone in solid state research, this group of superconductors has grown to the largest one. \cite{Kamihara-2008} The recently discovered compound \mbox{SrPtAs} exhibits superconductivity below $T_\mathrm{c}\approx 2.4~\mathrm{K}$ without doping, as found by Nishikubo et al. \cite{Nishikubo-2011} The transition metal pnictide superconductors contain layers of square lattices formed by the transition metal elements. In contrast to that, SrPtAs crystallizes in a hexagonal structure of weakly coupled non-centrosymmetric PtAs-layers, in which the charge transport takes place. Adjacent PtAs-layers are inverted to each other so that the bulk is centrosymmetric. Thus, SrPtAs is a prime example for staggered non-centrosymmetricity. \cite{Fischer2011} A well-known compound with a globally equivalent AlB$_2$-type structure, MgB$_2$, exhibits multigap superconductivity with $T_\mathrm{c}\approx40~\mathrm{K}$. \cite{souma-mgb2} Interestingly, very recent $\mu$SR experiments proved the development of a small static spontaneous internal field just below $T_\mathrm{c}$, evidencing time reversal symmetry (TRS) breaking in superconducting SrPtAs. \cite{Biswas-2012-muSR-prb} Different scenarios for this spontaneous TRS breaking are theoretically conceivable, all of which involve unconventional pairing states. Of these, the chiral $d$-wave superconducting state seems to provide the most consistent explanation of the experimental observations. This
superconducting state of \mbox{SrPtAs} is particularly exciting, as it hosts striking topological phenomena such as chiral Majorana surface states and bulk Majorana Weyl nodes. \cite{fischer13} Therefore, it is vital to understand the nature of the superconducting pairing in SrPtAs. In this context, nuclear magnetic resonance / nuclear quadrupole resonance (NMR/NQR) is one of the most powerful tools to shed light on such issues.

In this paper, we present $^{75}$As-NQR investigations to determine the superconducting properties of \mbox{SrPtAs}.  In the first stage of unconventional superconducting stateour experiments, $^{75}$As-NMR experiments at $40~\mathrm{MHz}$ were performed to extract the NQR frequency of SrPtAs in the metallic state, while a low upper critical field $H_{\mathrm{c2}}(0)$ of approximately $2000~\mathrm{Oe}$ does not allow NMR investigations in the superconducting phase. Therefore we carried out NQR experiments in a wide temperature range from $0.15$ to $15~\mathrm{K}$. NQR experiments can be performed in zero external static magnetic field. Since $^{75}$As is a spin $I=3/2$ nucleus, the NQR transitions ($\pm 1/2 \longleftrightarrow \pm 3/2$) result in a single line in the NQR spectrum. In addition to that, the nuclear magnetization recoveries are expected to exhibit a simple exponential form and a straightforward and unambiguous determination of the nuclear spin-lattice relaxation rate $1/T_1$ is possible. The temperature dependence of $1/T_1$ contains information about the symmetry of the superconducting order parameter.

Our main results are the evidence of multigap superconductivity with very weak inter-band coupling. The strongly suppressed Hebel-Slichter-Peak points to an unconventional superconducting state, such as the proposed chiral \emph{d}-wave or \emph{f}-wave order parameter. However a line node superconducting order parameter is not consistent with our data.

\section{Basic NMR and NQR properties}

Polycrystalline samples of SrPtAs were prepared via a solid state reaction method as described in Ref. \cite{Nishikubo-2011}. NMR/NQR experiments were carried out with a pulsed spectrometer in dilution fridge, He-3 and He-4 cryostats. $T_1$ was measured by monitoring the nuclear magnetization recovery after a saturation radio frequency (RF) pulse at a frequency of $f=27.75~\mathrm{MHz}$ both in the normal and superconducting state. Great care was taken in measuring 1/$T_1$ to avoid possible RF heating at very low temperatures. $T_\mathrm{c}$ of sample B is around 2.0 K, which has been estimated by \emph{in situ} AC susceptibility (ACS) measurements by using the NMR coil. The ACS measurements during cooling down and heating up the sample are plotted in the inset of Fig. \ref{fig:fig2}.

\begin{figure}
\includegraphics[width=\columnwidth]{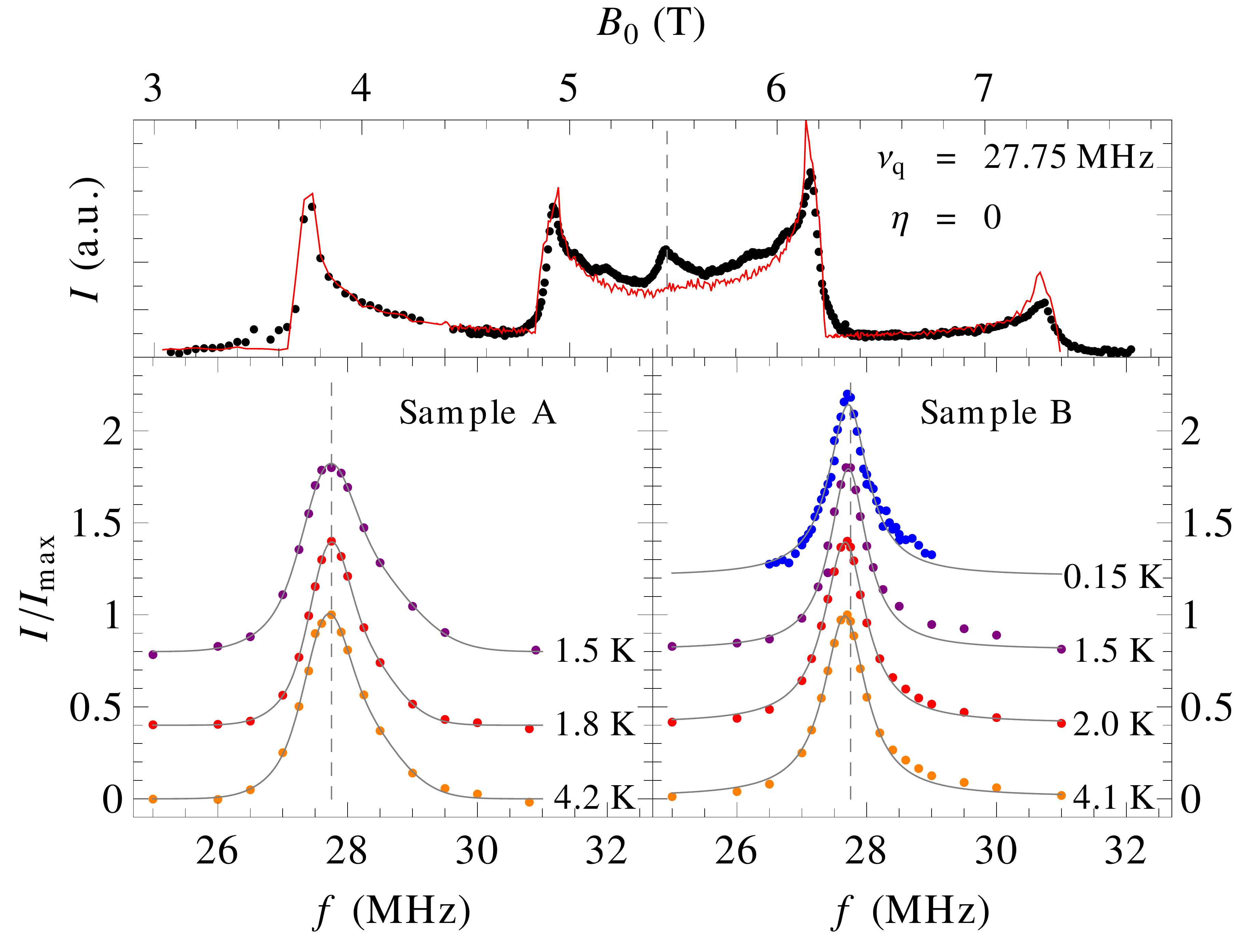}
\caption{\label{fig:fig1}Top: $^{75}$As-NMR field sweep spectrum of sample A, recorded at 13~K and 40~MHz (black dots) versus the simulation graph (red line) as described in the text. The dotted vertical line indicates the $^{75}$As-NMR larmor field. Left lower panel: $^{75}$As-NQR spectrum of sample A at three temperatures and double gaussian fit, right lower panel: $^{75}$As-NQR spectrum of sample B with a lorentzian fit, vertical dashed line: \mbox{$\nu_\mathrm{q}=27.75~\mathrm{MHz}$}.}\label{fig1}
\end{figure}

Figure~\ref{fig:fig1} (upper panel) shows the field sweep $^{75}$As NMR spectrum at a fixed frequency of $40~\mathrm{MHz}$ and \mbox{$T=13~\mathrm{K}$}. The spectrum represents a typical $I=3/2$ nucleus in case of strong quadrupole interaction. To determine the $^{75}$As-NQR frequency $\nu_\mathrm{q}$ and the asymmetry parameter $\eta$, we diagonalized the nuclear Hamiltonian
\begin{equation}
\label{eq:spectra-simulation}
H~=~-\gamma \hbar (\textbf{B}_\mathrm{ext}+\textbf{B}_\mathrm{hyp})\textbf{I}+h\frac{\nu_\mathrm{q}}{6}\left[3I_z^2-\textbf{I}^2+\eta\left(I_x^2-I_y^2\right)\right]
\end{equation}
at each sweep step for 25000 random orientations of the external field $\mathbf{B}_\mathrm{ext}$
with respect to the electric field gradient (EFG) principle axis $z$.
The best fitting quadrupole parameters are $\nu_\mathrm{q}=27.75~\mathrm{MHz}$, $\eta=0$
and no internal hyperfine field $\textbf{B}_\mathrm{hyp}=0$ was found.
Additional intensity located near the center at around $5.5~\mathrm{T}$  might be due to surface effects, parts of a different crystalline phase (SrPt$_2$As$_2$) or partial oxidation.

As shown in Fig.~\ref{fig:fig1} (lower panel), the $^{75}$As-NQR spectrum was observed around $27.75~\mathrm{MHz}$. Note that this value of the $^{75}$As-NQR frequency in SrPtAs is much larger than in other transition metal pnictides \cite{{CeFeAsPO-NMR-Rajib},{Rajib-EuFe2As2-NMR}}.
Due to a very small upper critical field of approximately $2000~\mathrm{Oe}$, the irradiating pulse disturbs superconductivity. This effect manifests itself in a shift of the coil inductance. Therefore, the equilibrium superconducting state is not present while pulsing. However, the superconducting properties recover very fast on the $\mathrm{\mu s}$ timescale, which is small compared to the nuclear relaxation times. This is verified by measurements of the shift of the coil inductance after a strong pulse and is in line with the field cycle experiment described in Ref. \cite{maclaughlin76}, where the pulses are applied in the normal state.

Figure~\ref{fig:fig1} (lower left panel, lower right panel) depicts several $^{75}$As-NQR spectra recorded in the normal and in the superconducting phase for the samples A and B. In the two regimes of the normal and superconducting state, the spectral shape does not change significantly, as well as $\nu_\mathrm{q}$ is constant over the whole temperature range. The single peak structure rules out the presence of any spurious phases in SrPtAs with a nearby $\nu_\mathrm{q}$. A homogeneous signal is further supported by the fact, that the $T_1$ measurements at different positions in the NQR spectrum give equal $T_1$ values.

Sample A and sample B are of different quality, as deduced from the NQR spectra. Sample A yields a linewidth (FWHM) of \mbox{$\approx 1~\mathrm{MHz}$}, while sample B has a linewidth of \mbox{$\approx700~\mathrm{kHz}$}. Both show a slight asymmetry that is largest at $1.5~\mathrm{K}$. This may be an effect of the crystal surface or impurities. 

\section{Nuclear relaxation rate}\label{sec:relax}

The nuclear magnetization recovery curves were fitted by
\begin{equation}
\label{eq:equation1}
m(t)= A \left(1- B~e^{-3\,t/T_1}\right),
\end{equation}
where $m(t)$ is the respective nuclear magnetization at a time $t$ after the saturation pulse, A and B are parameters that determine scaling and an offset. The recovery curves could be described well by this single exponential function except at low temperatures, where a stretched exponential function $m(t)= A (1-B\,\exp((-3\,t/T_1)^{\beta}) $ is required. The stretching parameter $\beta$ ranges from $\approx\!0.5$ at low temperatures $T<400~\mathrm{mK}$ to $\approx\!1$ at $T>1.2~\mathrm{K}$. This behavior may be an effect of the irradiating pulse. A rather large scattering of the stretching parameter ($\pm 0.1$) is found to be an experimental artefact that does not affect the $T_1$ observable in a perceptible amount. This is ensured by some measurements with varied experimental parameters (in particular pulse power and duration), that show no deviation of the spin-lattice relaxation times with respect to the confidence interval. These test measurements were done at several temperatures. As mentioned above, the possibility of multiple relaxation channels in the spin-lattice relaxation recovery process originating from a foreign structural phase is ruled out.

\begin{figure}
\includegraphics[width=\columnwidth]{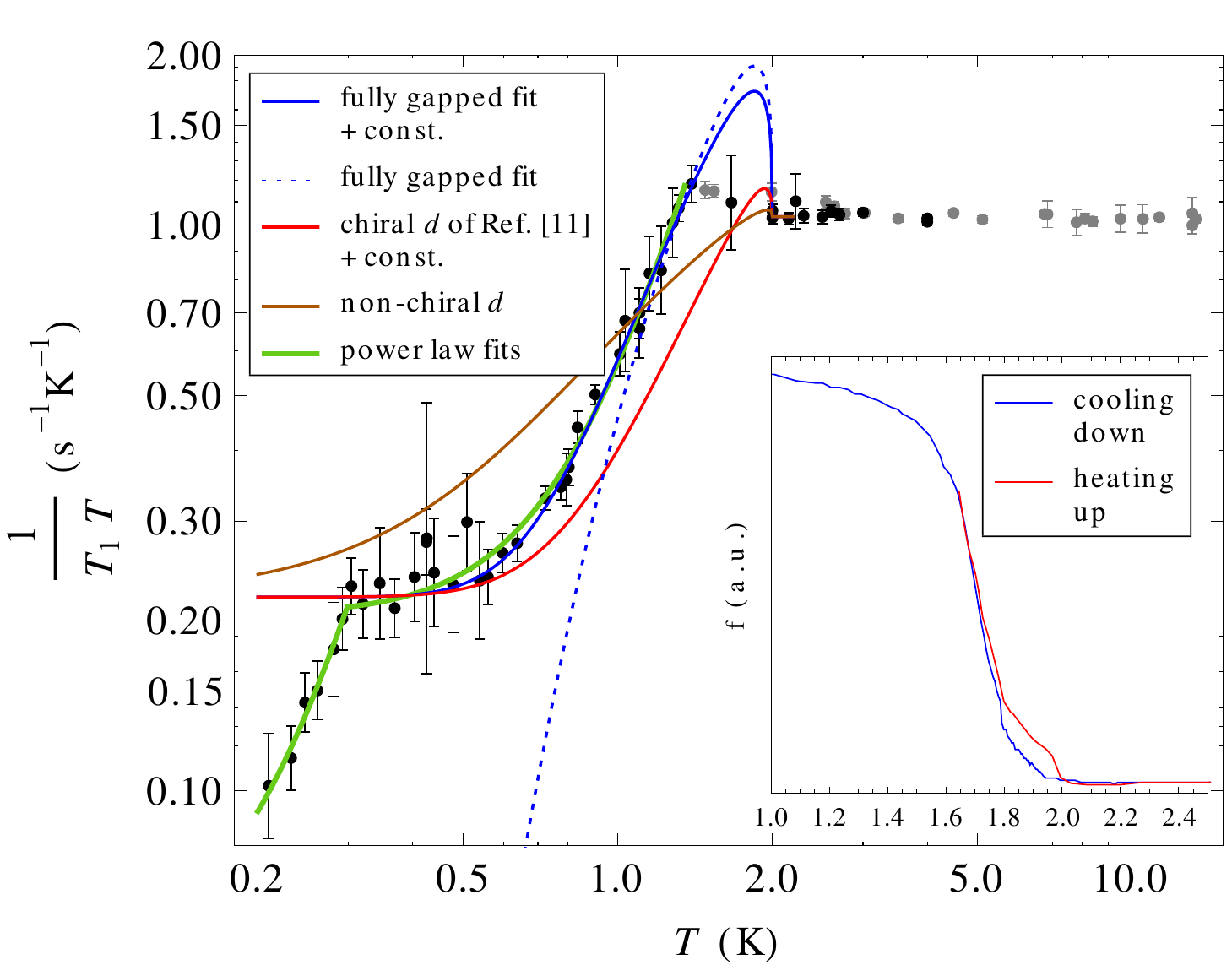}
\caption{\label{fig:fig2}$^{75}$As spin-lattice relaxation rate versus temperature for sample B (black) and sample A (gray) with fits and simulations (see text). The hump at $300~\mathrm{mK}$ indicates the superconducting transition of the second superconducting band. The fully gapped fit for the primarily superconducting band supports \emph{s}-wave, chiral \emph{d}-wave and \emph{f}-wave gap. A non-chiral \emph{d}-wave model does not fit the data. The inset shows the temperature dependency of AC susceptibility data at zero field using the \emph{in situ} NMR coil, revealing $T_\mathrm{c} \approx 2.0~\mathrm{K}$.}\label{fig2}
\end{figure}

In Fig. \ref{fig:fig2}, the results of our $^{75}$As NQR spin-lattice relaxation rate study are shown. While the presented data were taken at $27.75~\mathrm{MHz}$, data were also recorded at different frequencies to approve the constancy over the NQR spectrum. In the normal metallic state, $1/T_1T$ follows the simple Korringa relation ($T_1T = 0.97~\mathrm{s\,K}$), as expected for SrPtAs.

The determination of $T_\mathrm{c}$ via ACS experiment (see inset of Fig. \ref{fig:fig2}) reveals a critical temperature of $\approx 2.0~\mathrm{K}$, which is in contrast to earlier findings of $\approx 2.4~\mathrm{K}$. \citep{{Nishikubo-2011},{Biswas-2012-muSR-prb}} The discrepancy between $T_\mathrm{c}$ measured by means of ACS and the temperature, at which the spontaneous internal field developed in former $\mu$SR measurements, is not clear. Since it is the idem sample, it may be an effect of ageing but also may point to multiple phase transitions. $1/T_1T$ shows a sudden strong decrease below $1.3~\mathrm{K}$, which would in general be expected for an unconventional superconductor at $T_\mathrm{c}$. A possible Hebel-Slichter peak is strongly suppressed in comparison with conventional \emph{s}-wave superconductors. A second very distinct feature of the relaxation rate is a hump with a sharp edge at $300~\mathrm{mK}$. Below this edge, the relaxation rate obeys a power law decrease similar to the behavior near $1.3~\mathrm{K}$.

Earlier work on MgB$_2$ explains the suppression of a Hebel-Slichter peak with strong coupling effects or a possible quasiparticle broadening. \cite{Kotegawa-MgB2-Prl-2001} This is not necessarily adaptable to SrPtAs, in view of the smaller $T_\mathrm{c}$, the broader peak and the rather sharp edge at $1.3~\mathrm{K}$.
The hump at $300~\mathrm{mK}$ indicates a residual density of states (DOS) at the fermi level, which suddenly disappears at $300~\mathrm{mK}$. This is explainable if one assumes multiband superconductivity with weak inter-band scattering (see below), which is in line with the suggestion of Nishikubo et al. for SrPtAs \cite{Nishikubo-2011} based on the upward curvature of $H_{\mathrm{c2}} (T)$. No Hebel-Slichter peak can be identified at this hump, which is consistent with an unconventional gap symmetry or might originate from finite inter-band coupling.

The complex band structure of SrPtAs with three pairs of Fermi surfaces, the wealth of possible superconducting states and a corresponding large number of free parameters
poses a challenge to the theoretical calculation of $1/T_1$. We therefore resort to two fits of $1/T_1$ assuming that fully gapped pairing functions have been realized. This approach is partially justified by the $\mu$SR measurements that revealed that extended nodes of the order parameter are unlikely. \citep{Biswas-2012-muSR-prb} Both model functions are based on a generalized Hebel-Slichter formula (cf. \cite{maclaughlin76}):
\begin{eqnarray}
\frac{T_{1\mathrm{N}}}{T_{1\mathrm{S}}}=\frac{2}{k_\mathrm{B} T} \int_{0}^{\infty} (N_\mathrm{s}(E) N_\mathrm{s}(E') + M_\mathrm{s}(E) M_\mathrm{s}(E')) \nonumber \\  \times f(E) (1-f(E')) dE, \label{HSequ}
\end{eqnarray}
with the density of states $N_\mathrm{s}$, the so-called anomalous density of states $M_\mathrm{s}$
\begin{equation}
\begin{aligned}
N_\mathrm{s}(E)&=\Re \left[\int{P(a)\dfrac{E}{\sqrt{E^2-\Delta_0^2\vert1+a\vert^2}}da}\right]\\
M_\mathrm{s}(E)&=\Re \left[\int{P(a)\dfrac{\Delta_0(1+a)}{\sqrt{E^2-\Delta_0^2\vert1+a\vert^2}}da}\right],
\end{aligned}
\end{equation}
the Fermi function $f$ and $E'=E+\hbar \omega_{\mathrm{nuc}}$. $P(a)$ is the distribution of the anisotropic superconducting gap $\Delta(\mathbf{\Omega})=\Delta_0(1+a(\mathbf{\Omega}))$, where $\mathbf{\Omega}$ is the Fermi surface parametrization. $P(a)=\delta(a)$ describes the isotropic BCS case.

The temperature dependence of the superconducting gap is calculated by solving the BCS gap equation numerically for a specified $T_\mathrm{c}$ and then scaling the solution to obtain quotients $\Delta_0/k_\mathrm{B} T_\mathrm{c}$ that differ from the BCS-result.

The blue curve in Fig \ref{fig:fig2} is calculated with a simple model: $P(a)$ is assumed to be a rectangular function, that is finite in the range $[\Delta_{0}(1-\delta/2),\Delta_{0}(1+\delta/2)]$. We set $T_\mathrm{c} = 2.0~\mathrm{K}$, $\delta=0.35$ and $\frac{\Delta_{0}}{k_\mathrm{B} T_\mathrm{c}}=1.06 \cdot 1.76$, and add a constant baseline to allow for the proposed multigap character (continuous line). This model implies that inter-band scattering of electrons during the relaxation process is suppressed, which is motivated by a very low inter-band coupling. However, the opening of the second gap is not included here. Obviously, the pronounced Hebel-Slichter peak is not consistent with our data but the decrease is very well reproduced, in contrast to the simple fully gapped fit without the added baseline (dashed line), thus a strong indication for multigap superconductivity is found. A commonly used interpretation of this model is an anisotropic \emph{s}-wave gap. The high $\delta$ necessary to fit the data and the suppression of a Hebel-Slichter peak points to an anisotropic/unconventional state. Other fully gapped symmetries such as chiral \emph{d}-wave or \emph{f}-wave, which show both TRS breaking, can reproduce the blue curve very well when assuming a much smaller anisotropy parameter of $\delta = 0.02$ (not shown in Fig. \ref{fig:fig2}). In these cases the order parameter is complex and $\langle \Delta \rangle=0$, and therefore $M_\mathrm{s}=0$. Note that since SrPtAs has a complex band structure, a Hebel-Slichter peak, even if suppressed, does not imply an \emph{s}-wave order parameter.

It is interesting how theoretical predictions, in particular for the chiral \emph{d}-wave state \citep{fischer13,Goryo2012}, compare with the present NQR study. With the knowledge of the bandstructure and a simple representation of the chiral \emph{d}-wave state provided by Ref.~\cite{Goryo2012}, we calculated the spin-lattice relaxation rate without any fitting parameters. This is plotted in Fig. \ref{fig:fig2} (red curve). The model clearly underestimates the relaxation rates, while the height of the Hebel-Slichter peak does agree with the data. The deviation can originate from momentum- and temperature-dependences of the chiral \emph{d}-wave gap that differ from the infinitesimal coupling form of Ref. \cite{Goryo2012} and the predictions of the BCS gap equation. Hence, the observed deviation does not exclude the chiral \emph{d}-wave state.

The non-chiral \emph{d}-wave order parameter produces a much flatter curve (see Fig. \ref{fig:fig2}, brown curve) than our data and a rather instant decrease, which does not match to our findings. Since this behavior is due to line nodes in the gap function, we can generally exclude line nodes in the superconducting order parameter.

\begin{table}
\begin{center}
\begin{ruledtabular}
\begin{tabular}{cccccccc}
$T_{\mathrm{c2}}(\mathrm{mK})$ & $a_1$ & $b_1$ & $c_1$ & $a_2$ & $b_2$ & $c_2$\\
\hline
$298$&$4.3(1)$&$0.37(1)$&$0.20(1)$&$4.4(28)$&$8.9(266)$&$0.05(7)$\\
\end{tabular}
\end{ruledtabular}
\end{center}
\caption{Parameters of the power law fit to the spin-lattice relaxation rate}\label{powfit}
\end{table}

Finally we include a phenomenological fit with the function
\begin{equation}
\left(\frac{1}{T_1}\right)_{\mathrm{pow}}=\begin{cases} c_1 T+ b_1 T^{a_1} & T>T_{c2}\\
 c_2 T + b_2 T^{a_2} + b_1 T^{a_1} & T\leq T_{c2}  \end{cases}.
\end{equation}
The physical interpretation of this function is a metallic two-band superconductor with separated bands (see above). This only makes sense in the limit of weak coupling between the bands. Results are summarized in \mbox{Tab. \ref{powfit}}. The main purpose is to provide a reasonable determination of the slope of $1/T_1T$, and the temperature $T_{\mathrm{c2}}$, where the second gap opens. The relaxation rate driven by the corresponding band is approximately $1/4$ as high as the relaxation driven by the primary superconducting band, which is derived from the parameter $c_1$. Since the relaxation rate is related to the DOS at the Fermi level, this would agree to the assumption that the Fermi surfaces that primarily contribute to superconductivity are the ones around the H- and K-point of the Brillouin zone, which host the majority of the DOS ($\approx 60\%$), and the second is one or both of the Fermi Surfaces around the $\Gamma$-Point.~\cite{Goryo2012} This is consistent with the mechanism proposed for chiral $d$-wave superconductivity: It is driven by superconducting fluctuations near the Van Hove singularities at the M-points which are very close in energy to that Fermi surface.

\section{Conclusion}

Detailed $^{75}$As-NQR investigations on two different polycrystalline samples of the locally non-centrosymmetric superconducting SrPtAs with $T_\mathrm{c} \approx 2.0~\mathrm{K}$ are presented. In the normal state, $1/T_1T = const.$ is found as expected for metallic SrPtAs. Below $T_\mathrm{c}$, the absent Hebel-Slichter coherence peak indicates unconventional superconductivity. The spin-lattice relaxation rate $1/T_1T$ decreases below $T_\mathrm{c}$ and levels off a finite value, which proves multigap superconductivity with very weak inter-band coupling. Eventually, the relaxation rate declines again just below $T_\mathrm{c2}=300~\mathrm{mK}$, due to the opening of the second gap. Several fits and simulations were carried out to describe the $1/T_1T$ data. A fully gapped model fits to the observed decreasing behavior. In establishing the multigap character of superconductivity in SrPtAs, our data provides guidance to revisit the theoretical models of the superconducting gap function, and constrains the possible pairing mechanisms.

\begin{acknowledgments}

The presented experiments were performed at the Technische Universit\"at Dresden, Germany, and at the National High Magnetic Field Laboratory USA, which is supported by National Science Foundation Cooperative Agreement No. DMR-1157490, the State of Florida, and the U.S. Department of Energy. Both samples were synthesized at the Ludwig-Maximilians-University Munich. We appreciate the support by the Deutsche Forschungsgemeinschaft through SA 2426/1-1, KU 3066/1-1 and the Research Training Group GRK 1621 at Technische Universit\"at Dresden.

\end{acknowledgments}

\nocite{*}
\bibliography{srptas-nmr}

\end{document}